\documentclass[preprint, aps, prl, amsmath, superscriptaddress, floatfix]{revtex4-1}

\usepackage[hidelinks,
    pdfauthor={C. Arnold, O. Vendrell, R. Welsch, R. Santra},
    pdftitle={Towards attochemistry: Control of nuclear motion through conical intersections and electronic coherences}
]{hyperref}

\usepackage{datetime}

\usepackage{amsmath}

\usepackage{units}
\usepackage{braket}

\usepackage{graphicx}
\usepackage[caption=false]{subfig}
\graphicspath{ {/} {./graphics/} {./graphics/plots/} }

\begin{document}

\title{Towards attochemistry: Control of nuclear motion through conical intersections and electronic coherences}

\author{Caroline Arnold}
\email{caroline.arnold@cfel.de}
\affiliation{Center for Free-Electron Laser Science, DESY, Notkestrasse 85, 22607 Hamburg, Germany}
\affiliation{Department of Physics, University of Hamburg, Jungiusstrasse 9, 20355 Hamburg, Germany}
\affiliation{The Hamburg Centre for Ultrafast Imaging, Luruper Chaussee 149, 22761 Hamburg, Germany}

\author{Oriol Vendrell}
\email{oriol.vendrell@phys.au.dk}
\affiliation{Center for Free-Electron Laser Science, DESY, Notkestrasse 85, 22607 Hamburg, Germany}
\affiliation{The Hamburg Centre for Ultrafast Imaging, Luruper Chaussee 149, 22761 Hamburg, Germany}
\affiliation{Department of Physics and Astronomy, Aarhus University, Ny Munkegade 120, 8000 Aarhus, Denmark}

\author{Ralph Welsch}
\email{ralph.welsch@cfel.de}
\affiliation{Center for Free-Electron Laser Science, DESY, Notkestrasse 85, 22607 Hamburg, Germany}

\author{Robin Santra}
\affiliation{Center for Free-Electron Laser Science, DESY, Notkestrasse 85, 22607 Hamburg, Germany}
\affiliation{Department of Physics, University of Hamburg, Jungiusstrasse 9, 20355 Hamburg, Germany}
\affiliation{The Hamburg Centre for Ultrafast Imaging, Luruper Chaussee 149, 22761 Hamburg, Germany}

\date{\today}

\begin{abstract}
The effect of nuclear dynamics and conical intersections on electronic coherences is investigated employing a two-state, two-mode linear vibronic coupling model. Exact quantum dynamical calculations are performed using the multi-configuration time-dependent Hartree method (MCTDH). It is found that the presence of a non-adiabatic coupling close to the Franck-Condon point can preserve electronic coherence to some extent. Additionally, the possibility of steering the nuclear wavepackets by imprinting a relative phase between the electronic states during the photoionization process is discussed. It is found that the steering of nuclear wavepackets is possible given that a coherent electronic wavepacket embodying the phase difference passes through a conical intersection. A conical intersection close to the Franck-Condon point is thus a necessary prerequisite for control, providing a clear path towards attochemistry. 
\end{abstract}

\maketitle


Ultrashort laser pulses allow to resolve electronic and nuclear motion in molecules on their natural timescales \cite{cor07, kra09, cer15a, cal16}. With the dawn of attosecond pulses, it is now possible to create coherent superpositions of excited electronic states of a photo-ionized molecule. Electronic coherences are believed to be important for a wide range of processes, e.g., electron hole oscillations \cite{cal14} and efficient energy conversion in light-harvesting complexes \cite{eng07}. 
In theoretical descriptions of electronic coherence, the nuclei are often fixed as they are heavy compared to the electrons. Such calculations predict long-lived coherences and electron hole migrations driven by electron correlation \cite{cal14, kul07, gol15, lin17}. However, recent quantum-dynamical studies show that the motion of nuclei cannot be neglected and that nuclear motion can lead to electronic decoherence within few femtoseconds \cite{hal13, her14, vac15a, pau16, vac17, arn17}. 

The interplay of electronic and nuclear motion becomes especially relevant in the presence of strong non-adiabatic couplings, as the Born-Oppenheimer separation breaks down and the timescales of electronic and nuclear motion become comparable \cite{sto13}. Non-adiabatic couplings are particularly strong at conical intersections (C.I.), which are abundant in the potential energy landscape of poly-atomic molecules \cite{dom04, kow15a}. First insight into the influence of C.I.s on electronic coherence was obtained recently with a quantum-dynamical treatment of paraxylene and BMA[5,5], but a systematic understanding remains elusive \cite{vac17}. 

Non-adiabatic couplings and C.I.s are already exploited in control schemes employing femtosecond laser pulses. The underlying processes are typically well-understood and the nuclear wavepacket can be steered to desired reaction products \cite{zew00, abe05, hof12a, sto13, lie16}. With attosecond pulses, due to their large width in the energy domain, it becomes feasible to control the electronic rather than the nuclear degrees of freedom. Through non-adiabatic couplings, the relative weight and phase between electronic states may affect the velocity as well as the direction of nuclear dynamics, as investigated in models of toluene and benzene employing approximate Ehrenfest dynamics \cite{vac15c, mei15}. This might open the path towards attochemistry, where, by controlling the relative phase between electronic states, nuclear dynamics on a time scale of tens to hundreds of femtoseconds is influenced \cite{sal12, lep14, nis17}. Thus, attochemistry will allow for directing the system towards desired, but unlikely reaction products. 

In this Letter, we present a systematic study of the influence of non-adiabatic couplings on electronic coherence and discuss possible pathways towards attochemistry by imprinting a relative phase between the electronic states forming a coherent superposition. To this end, we employ a two-state, two-mode model system and consider different positions of the C.I. relative to the Franck-Condon region \cite{wor04} as well as different coupling strengths and relative phases.


The linear vibronic coupling Hamiltonian \cite{kop84} is employed to describe the potential energy surfaces of two electronically excited states in a local diabatic picture. Two coordinates forming a Jahn-Teller type C.I., the tuning mode $x$ and the coupling mode $y$, are considered in mass- and frequency-weighted ground-state normal modes. The corresponding excited-state Hamiltonian reads 
\begin{equation}
    H = 
    \begin{pmatrix}
        T + V_1(x, y)   &  W_{12} \\
        W_{12}   &   T + V_2(x, y) + \Delta E
    \end{pmatrix}
    ,
    \label{eq:lvcm-hamiltonian}
\end{equation}
with the kinetic energy operator $ T $ and the two diabatic states given as
$
V_{1,2}(x, y) 
=
\frac{\gamma}{2} \left(x^2 + y^2\right) 
+ 
\kappa_{1,2}^{(x)} x 
+ 
\kappa_{1,2}^{(y)} y,
$
where $\gamma$ refers to the vibrational frequencies of the excited state, $\kappa^{(x,y)}_{1,2}$ defines the slope at the C.I. along $x$ and $y$, and $\Delta E$ is the gap at the Franck-Condon point $(x_\mathrm{C.I.} = y_\mathrm{C.I.} = 0)$. The non-adiabatic coupling is introduced by $W_{12}=\lambda y$. It is considered up to first order and its strength is varied between $\lambda=\unit[0.0]{a.u.}$ and $\unit[0.02]{a.u.}$ Throughout this letter, atomic units (a.u.) are used. The C.I. is moved to arbitrary positions $(x_\mathrm{C.I.}, y_\mathrm{C.I.})$ by adjusting the model parameters. Details on the model and the numerical parameters can be found in the supplemental material (S.M.) \cite{supplement}. 

The initial state is assumed to be an equally weighted coherent superposition of both electronic states, where the ground-state nuclear wavepacket is lifted vertically to the diabatic potential energy surfaces, thus modeling a short-time impulsive excitation from the common electronic ground state to the excited-state manifold: 
\begin{equation}
    \braket{x,y | \Psi(t=0)} = c_1 \chi_1(x,y) \ket{1} + c_2 \chi_2(x,y) \mathrm{e}^{i \varphi} \ket{2},
    \label{eq:initial-state-superposition}
\end{equation}
where $c_1 = c_2 = 1 / \sqrt 2$, $\chi_1 = \chi_2 = \chi$, and $\varphi$ is a relative phase between the electronic states. The ground-state nuclear wavepacket is given as a product of Gaussians,
\begin{equation}
    \chi(x,y) = \frac{1}{\sqrt \pi} \mathrm{e}^{-(x^2+y^2)/2}.
    \label{eq:initial-state-nuclear-wavepacket}
\end{equation}
The wavepacket is propagated employing the Multi-Configuration Time-Dependent Hartree method (MCTDH) in its multiset implementation in the Heidelberg package \cite{mctdhMLpackage, bec00, mctdh1990}. The numerical accuracy of the simulations is assured by adjusting the number of single-particle functions (SPF) used such that the natural weight of the highest SPF is below $10^{-4}$ \cite{bec00}. 

A basis-independent measure for the electronic coherence is given by the electronic purity $\mathrm{Tr}(\rho^2)$ \cite{arn17, vac17}, where $\rho$ is the reduced density matrix of the electronic subsystem expressed as
\begin{equation}
    \rho_{\mu\nu}(t) = \int \mathrm{d}x\,\int\mathrm{d}y\, \braket{\mu | \Psi(t)} \braket{\Psi(t) | \nu}.
    \label{eq:reduced-density-matrix-element}
\end{equation}
For our two-state system, $\mathrm{Tr}(\rho^2) = \rho_{11}^2 + \rho_{22}^2 + 2 |\rho_{12}|^2$. Note that $\mathrm{Tr}(\rho^2) = 1$ corresponds to a fully coherent electronic superposition, and, for $c_1 = c_2$, $\mathrm{Tr}(\rho^2) = 0.5$ to an incoherent mixture. Electronic decoherence is caused through three mechanisms \cite{fie03, vac17}: \textit{(i)} dephasing due to the width of the nuclear wavepacket, \textit{(ii)} loss of overlap of nuclear wavepackets propagated on different potential energy surfaces, and \textit{(iii)} transfer of nuclear density between electronic states. From an analytic expansion of electronic density matrix elements up to second order in time, and considering the initial state given in Eq.~\eqref{eq:initial-state-superposition}, it can be shown that the diabatic populations $\rho_{11}, \rho_{22}$ are constant up to second order in time, while the coherences are phase-dependent in the presence of a non-adiabatic coupling $\lambda$:
\begin{align}
    |\rho_{12}|^2
    =
    &
    \,
    |c_1|^2 |c_2|^2
    -
    t^2 |c_1|^2|c_2|^2 \braket{\chi | (H_1 + H_2)^2 | \chi}
    \nonumber \\
    &
    - 2 t^2 \lambda^2 |c_1|^2|c_2|^2 \braket{\chi | y^2 | \chi} \sin^2 \varphi
    +
    \mathcal{O}(t^3)
    ,
    \label{eq:rho12}
\end{align}
where $H_\mu = T + V_\mu + \Delta E_\mu, \mu = 1,2$. The second term in Eq.~\eqref{eq:rho12} is due to decoherence caused by the dephasing and loss of overlap (mechanisms \textit{(i)} and \textit{(ii)}) while the third term is due to the coupling of the electronic states. The latter is the only term carrying a phase dependence to second order. Hence, the influence of non-adiabatic coupling on decoherence can be controlled by the relative electronic phase. It vanishes in second order for the case of $\varphi = 0$, i.e., no relative phase is imprinted on the electronic states. Details can be found in the S.M.


\begin{figure*}
    \centering
    \includegraphics{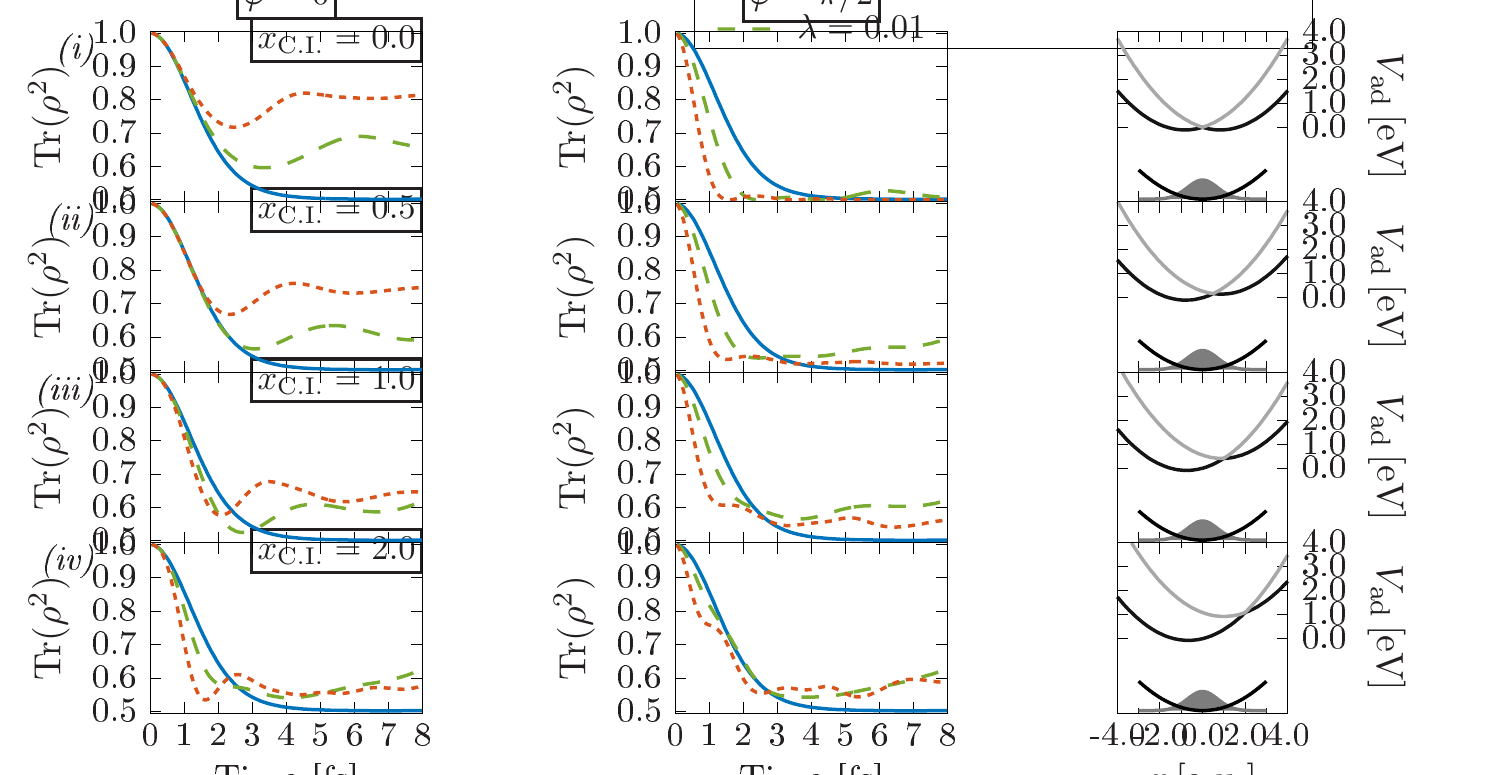}
    \caption{
        Time-dependent electronic purity (see Eq.~\eqref{eq:reduced-density-matrix-element}) of an initially equally weighted superposition of two electronic states in the presence of a C.I. for different values of the non-adiabatic coupling strength $\lambda$ and no relative phase (left, $\varphi = 0$, see Eq.~\eqref{eq:initial-state-superposition}) and $\varphi = \frac \pi 2$ (center). On the right hand side, cuts through the adiabatic potential energy surfaces at $y=0$ are shown. The ground state is included as well as the initial wavepacket. From top to bottom, the C.I. is \textit{(i)} at the Franck-Condon point, \textit{(ii)} within, \textit{(iii)} at the edge of, and \textit{(iv)} outside the Franck-Condon region. 
    }
    \label{fig:trace_squared_xci}
\end{figure*}


The electronic purity for different positions of the C.I. along the tuning mode, for different relative electronic phases and coupling strengths, is shown in Fig.~\ref{fig:trace_squared_xci}. In the adiabatic case $(\lambda = \unit[0]{a.u.})$, dephasing and the loss of spatial overlap between the nuclear wavepackets evolving on the different potential energy surfaces leads to ultrafast electronic decoherence within a few femtoseconds, in accordance with the results obtained with adiabatic models \cite{vac15a, arn17}. In the presence of non-adiabatic couplings, the relative electronic phase affects the electronic purity. For $\varphi = 0$, if the C.I. is located within the Franck-Condon region, defined with respect to the initial extension of the ground-state nuclear wavepacket, the coupling region is reached before decoherence occurs. In these cases, the non-adiabatic coupling preserves coherence to some extent, see panels~\textit{(i)}--\textit{(iii)}. However, if the intersection is located far from the Franck-Condon point as in panel~\textit{(iv)}, decoherence takes place before the intersection is reached. By imprinting a relative electronic phase of $\varphi = \frac \pi 2$, electronic coherences are destroyed rapidly even for a strong non-adiabatic coupling and a C.I. within the Franck-Condon region. This is in accordance with the analytic result at short times given in Eq.~\eqref{eq:rho12}.  In all cases, we do not observe a substantial increase in coherence upon the passage of the wavepacket through the C.I. \cite{lie16}. The full electronic purity can be decomposed into different contributions related to the three decoherence mechanisms. For all cases considered here, dephasing (mechanism \textit{(i)}), is the main cause of decoherence. Contributions from mechanism \textit{(iii)} are small, but get stronger the further the Franck-Condon point is located from the C.I. The dynamical evolution of the wavepackets causes the small revivals seen in the full electronic purity. Details can be found in the S.M. We also performed similar calculations for shifts of the C.I. along the coupling mode $y$. In this case, for strong non-adiabatic coupling the initial superposition reduces to the trivial case of a pure state involving only one adiabatic surface. The corresponding results can be found in the S.M.

In recent work, the electronic decoherence mechanisms in two specific molecules were studied in a non-adiabatic, quantum-dynamical framework \cite{vac17}. It was seen that the decoherence time in paraxylene with a C.I. near the Franck-Condon point amounts to $\unit[3]{fs}$, while in BMA[5,5] with a C.I. far from the Franck-Condon point, it amounts to $\unit[6]{fs}$. As was pointed out in Ref.~\cite{vac17}, the decoherence time is due to a complex interplay of several mechanisms influenced by different molecular parameters. With our model system we can disentangle the contributions of the position of the C.I. or the non-adiabatic coupling, while keeping all other PES parameters the same, which is not possible if specific molecules are used. We find that, for $\varphi = 0$, the further the C.I. is from the Franck-Condon point, the faster the decoherence, and the stronger the non-adiabatic coupling, the more coherence can be preserved.


\begin{figure}
    \centering
    \includegraphics{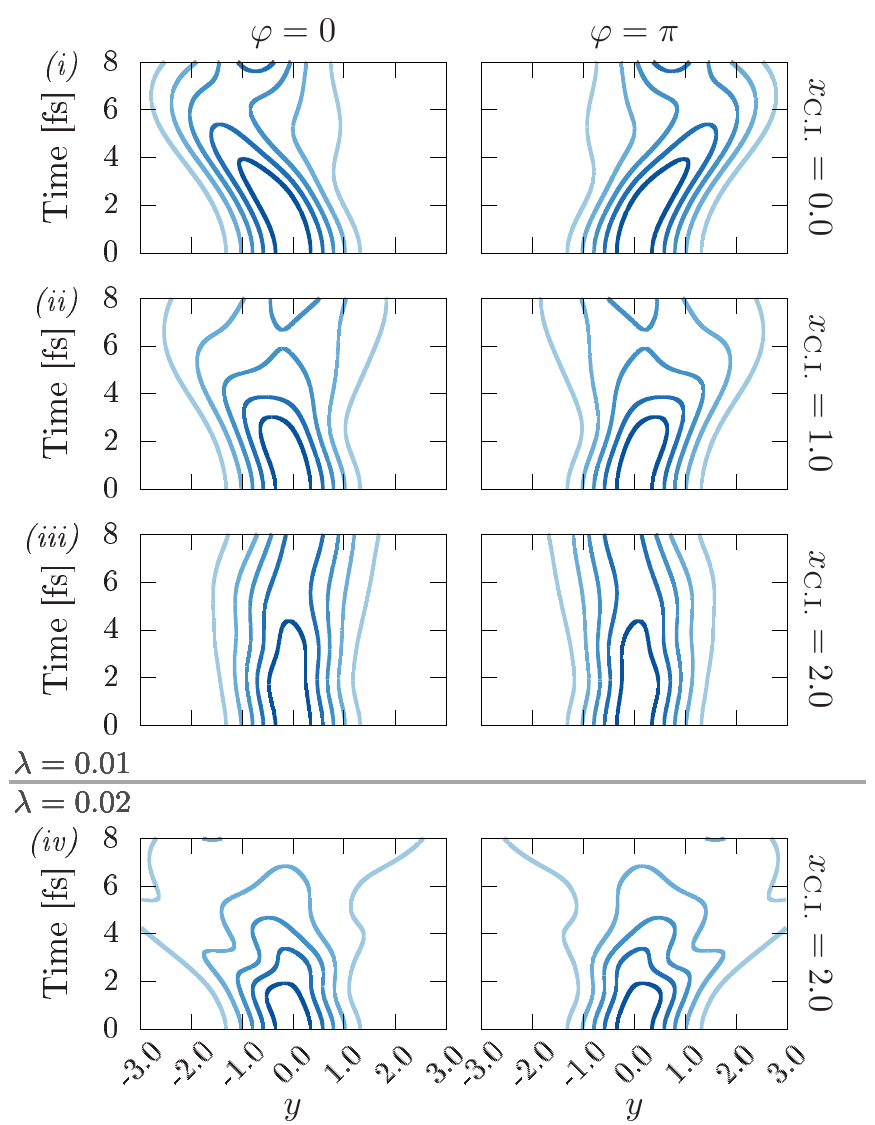}
    \caption{One-dimensional nuclear density along the coupling coordinate $y$ for different relative phases $\varphi$ imprinted on the electronic states and non-adiabatic coupling $\lambda =  \unit[0.01]{a.u.}$, \textit{(i)--(iii)}, and stronger non-adiabatic coupling,  $\lambda =  \unit[0.02]{a.u.}$, \textit{(iv)}, respectively. The C.I. is \textit{(i)} at the Franck-Condon point, \textit{(ii)} at the edge of, and \textit{(iii)--(iv)} outside the Franck-Condon region. 
    }
    \label{fig:ev_q2_xci_lambda_01_02}
\end{figure}

\begin{figure}
    \centering
    \includegraphics[width=.5\textwidth]{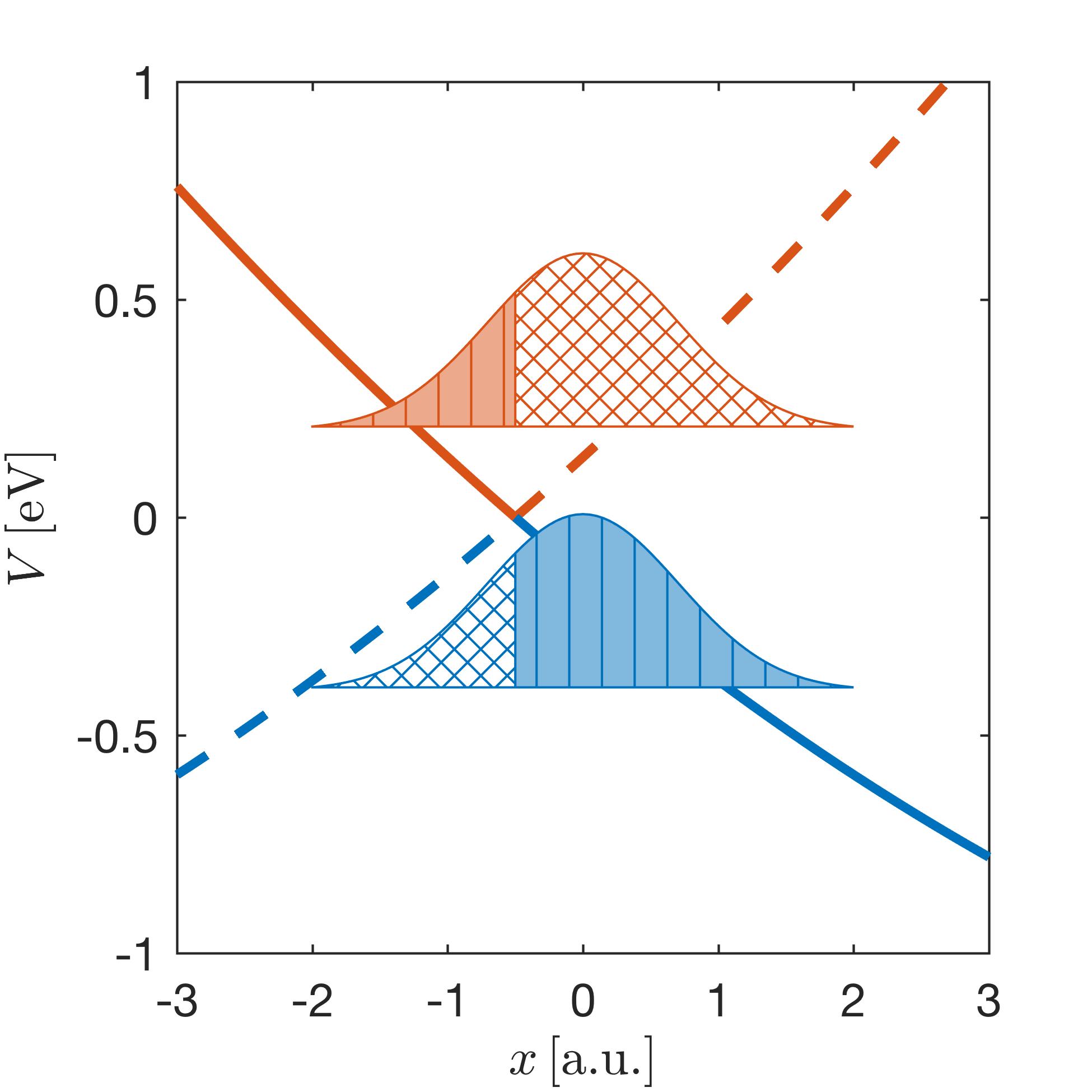}
    \caption{%
        Schematic of the nuclear wavepackets on the two adiabatic surfaces \textit{(red and blue)}. The wavepackets are created as a coherent superposition of ground-state wavepackets on two diabatic surfaces \textit{(dashed lines)} including an electronic phase difference \textit{(shading)}. Here, the wavepacket extends across the C.I., and thus the projection on the upper adiabatic surface \textit{(red)} embodies the phase difference. This leads to interference when the wavepacket moves towards the C.I. The part projected on the lower adiabatic surface \textit{(blue)} embodies the phase difference as well, but moves away from the C.I. and is not directly relevant for control. 
    }
    \label{fig:wavepackets_coin_sketch}
\end{figure}

The $\varphi$-dependence of time-dependent expectation values of nuclear coordinates is a potential path to attochemistry: a nuclear wavepacket could be steered in the desired direction by imprinting a relative phase between electronic states. For the system described by the Hamiltonian in Eq.~\eqref{eq:lvcm-hamiltonian} and the initial wavefunction given in Eq.~\eqref{eq:initial-state-superposition}, the expectation value of the tuning coordinate $y$ is expanded up to second order in time as
\begin{equation}
    \langle y \rangle (t)
    =
    \langle y \rangle_1 (t)
    +
    \langle y \rangle_2 (t)
    - 
    t^2 \lambda c_1 c_2 \cos \varphi
    +
    \mathcal{O}(t^3)
    ,
    \label{eq:y_expectation_value}
\end{equation}
where 
$
\langle y \rangle_\mu
=
|c_\mu|^2 \braket{ 
	\chi_\mu
	|
	y + t[H_\mu, y] + t^2[[H_\mu,y],H_\mu]
	|
	\chi_\mu
}
$
is the motion of the nuclear wavepackets on the uncoupled diabatic state $\ket{\mu}$ and $H_\mu = T + V_\mu + \Delta E_\mu$. The uncoupled motion of the nuclear wavepackets on the diabatic states is modified by the non-adiabatic coupling at second order in time. This modification also carries a phase dependence which allows for the steering of the nuclear dynamics by controlling the electronic phase. Note that the relative electronic phase between electronic states that are not coupled is irrelevant for the motion of the nuclei. Within the model considered here, $\langle x \rangle(t)$ is independent of the relative phase. Details of the derivation and the expectation value of an arbitrary chemical observable can be found in the S.M. 

In Fig.~\ref{fig:ev_q2_xci_lambda_01_02}, panels~\textit{(i)--(iii)}, we present the time-evolution of the one-dimensional density along the coupling coordinate $y$ for non-adiabatic coupling $\lambda = \unit[0.01]{a.u.}$ and the different positions of the C.I. employed before. If the electronic coherence persists once the nuclear wavepacket reaches the region of non-adiabatic coupling, then it can be steered along $y$ by varying $\varphi$, see panels~\textit{(i)}--\textit{(ii)}. Once electronic coherence is lost, the wavepacket cannot be controlled, see panel~\textit{(iii)} for a C.I. far from the Franck-Condon region. At the same C.I. position and with strong non-adiabatic coupling $(\lambda = \unit[0.02]{a.u.})$, control can be achieved even in this setting, see panel~\textit{(iv)}. This implies that nuclear controllability requires the possibility of interference, at the C.I., of the wavepackets initially created on different diabatic surfaces, carrying a phase difference, as indicated in Fig.~\ref{fig:wavepackets_coin_sketch}. Electronic decoherence suppresses this. This view is further validated by considering the evolution of the part of the wavepacket that is projected on the upper adiabatic potential energy surface. In this case, control is still possible, if the C.I. is close to the Franck-Condon point or for strong non-adiabatic coupling (see S.M.). If the C.I. is close to the Franck-Condon point and thus the energy separation between the electronic states is small and the electron dynamics is on a femtosecond rather than an attosecond time scale \cite{sto13}, coherent superpositions might be created by femtosecond pulses. For cases of strong non-adiabatic coupling and excitations far away from the C.I., the separation of the electronic states becomes larger and the use of broadband attosecond pulses is required for the excitations, leading to ''true'' attochemistry. 

Creating a coherent initial state with an imprinted phase by ultrashort pulses is an experimental challenge. Beyond the limit of sudden ionization employed in this work, nuclear dynamics and entanglement with the photoelectron may decrease the degree of initial electronic coherence reached in the remaining cation \cite{cal16a, nis17}. Coherent two-color pulses can in principle be used to excite two electronic states with varying relative phases \cite{kra09}. To find the optimal pulses, methods from coherent control of quantum phenomena or quantum optical control could then be adapted \cite{sha03, rab00, hof12, kli13}. Light-induced C.I., created in molecules with the help of external laser fields, could be used to control the position of the intersection and the strength of the non-adiabatic coupling \cite{moi08, nat16}. 


To conclude, we discussed the influence of non-adiabatic dynamics and relative electronic phases on electronic coherences created by ultrashort pulses. It is found that non-adiabatic coupling stabilizes electronic coherences if the C.I. is close to the Franck-Condon point. The further the C.I. is from the Franck-Condon point, the stronger the decoherence. Changing the relative electronic phase may enhance decoherence. If the wavepacket maintains electronic coherence in the region of the C.I., it can be steered in a desired direction by a relative phase imprinted initially between the electronic states. This steering of nuclear wavepackets opens a clear, but limited, path towards attochemistry. While attochemistry will not create new reaction pathways, it will provide steering possibilities along less likely paths. Novel schemes can then be developed to follow light-induced chemical reactions on an attosecond timescale and to control chemical observables by manipulating the electronic degrees of freedom.


\begin{acknowledgments}
This work has been supported by the excellence cluster \textit{The Hamburg Centre for Ultrafast Imaging - Structure, Dynamics and Control of Matter at the Atomic Scale} of the Deutsche Forschungsgemeinschaft.
\end{acknowledgments}




%
\end{document}